# ANALYTIC EXPRESSIONS OF HYDROTHERMAL WAVES [*]

Robert Conte
Service de physique de l'état condensé
CEA–Saclay
F–91191 Gif-sur-Yvette Cedex, France
Conte@drecam.saclay.cea.fr

Micheline Musette
Dienst Theoretische Natuurkunde
Vrije Universiteit Brussel
Pleinlaan 2
B–1050 Brussels, Belgium
MMusette@vub.ac.be, 27 October 1999

When subjected to a horizontal temperature difference, a fluid layer with a free surface becomes unstable and hydrothermal waves develop in the bulk. Such a system is modelized by two coupled amplitude equations of the one-dimensional, complex, cubic Ginzburg-Landau type. By transposing the method developed for one CGL3 equation, we obtain several new exact solutions expressed by closed form, singlevalued, analytic expressions. Some of them are the analogue of the famous amplitude hole solution of Bekki and Nozaki.

## 1. Introduction

Consider a thin fluid layer with a free surface, subjected to a horizontal difference of temperature between its vertical edges. When the temperature difference $T_2 - T_1$ is small, the basic flow is just one long, narrow convection cell. If one increases $T_2 - T_1$, this basic flow becomes unstable *via* a supercritical Hopf bifurcation, and traveling hydrothermal waves appear [1, 2]. Two waves develop (see Fig. 1 in Ref. [2]), and their complex amplitudes $(A_+, A_-)$ evolve in time according to [3]

$$\partial_t A_\pm = rA_\pm \mp v\partial_x A_\pm + (1+i\alpha)\Delta A_\pm - (1+i\beta)(|A_\pm|^2 + \gamma|A_\mp|^2)A_\pm, \quad (1)$$

in which $r$ is real and proportional to $T_2 - T_1$, while $(v, \alpha, \beta, \gamma)$ are real parameters and $\Delta$ is the Laplacian.

These are two coupled nonlinear partial differential equations of a very generic type, namely the cubic complex Ginzburg-Landau one (CGL3).

## 2. The problem for two coupled one-dimensional CGL3 equations

Let us for simplification consider one space dimension.
The scalar one-dimensional CGL3 equation for $A(x,t)$ is defined as

$$E \equiv iA_t + pA_{xx} + q|A|^2 A - i\gamma A = 0, \ pq \neq 0, \ (A,p,q) \in \mathcal{C}, \ \gamma \in \mathcal{R}, \qquad (2)$$

with $p, q, \gamma$ constant. The review [4] recalls the physical phenomena (pattern formation, superconductivity, nonlinear optics, ...) modelized by this equation and summarizes the known exact results.

Two coupled one-dimensional CGL3 equations are similarly defined as

$$E \equiv \begin{cases} iA_t + ivA_x + pA_{xx} + q(|A|^2 + \delta|B|^2)A - i\gamma A = 0, \\ iB_t - ivB_x + pB_{xx} + q(|B|^2 + \delta|A|^2)B - i\gamma B = 0, \end{cases} \qquad (3)$$

in which the coupling parameter $\delta$ is a complex constant and $v$ is the group velocity. These describe for instance the amplitudes of two lasers [5], or the spatiotemporal intermittency [6, 7], or hydrothermal waves [2].

The problem which we address is to find *exact solutions*, i. e. analytic expressions written in closed form, able to describe the observed topological structures : pulses, fronts, holes, ... The closed form which we require excludes any perturbation process, which would involve infinite series generically unable to *globally* represent the solution.

More specifically, we restrict to the search for the most general solitary wave

$$A(x,t) = a(\xi)e^{i(\omega_A t + \varphi_A(\xi))}, \ B(x,t) = b(\xi)e^{i(\omega_B t + \varphi_B(\xi))}, \ \xi = x - ct, \qquad (4)$$

in which $(a, b, \varphi_A, \varphi_B)$ are functions of the reduced independent variable $\xi$, and we only consider the strict CGL3 case $\text{Im}(p/q) \neq 0$.

## 3. Analyticity restricts the number of free constants

The chaotic nature of CGL3 restricts the number of arbitrary constants the solitary wave (4) may depend on. Let us explain how on the Kuramoto-Sivashinsky equation, for which the solitary wave reduction is the ordinary differential equation (ODE)

$$\nu u''' + \mu u' + u^2/2 + K = 0, \ (\mu, \nu) \text{ real constants}, \qquad (5)$$

in which $K$ is an integration constant. Denoting $\chi = x - x_0$, the general solution is locally represented as [8, 9]

$$u(x_0, c_+, c_-) = 120\nu\chi^{-3}\{\text{Taylor}(\chi) \\ + c_+\chi^{(13+i\sqrt{71})/2}(\dots) + c_-\chi^{(13-i\sqrt{71})/2}(\dots)\}, (6)$$

in which $(x_0, c_+, c_-)$ are arbitrary constants, Taylor a converging Taylor series of $\chi$, and the two $(\dots)$ highly multivalued series depending on $\chi$ with only positive integral powers of $(c_+, c_-)$. The presence of chaos is linked to the fact that the series (6) is what Painlevé calls "une solution illusoire", and the only way to possibly recover analyticity is to require $c_+ = c_- = 0$, restricting the illusory general solution to what can be called the *general analytic solution*, depending on $3 - 2 = 1$ arbitrary constant, namely $x_0$.

The above two irrational exponents are computed as follows [10], in the example of the ordinary differential equation (5).

In a first step, one looks for a singular dominant behaviour $u \sim u_0 \chi^p$ (with $u_0$ nonzero and $p$ not a positive integer) when $x \to x_0$. The term $u'$ is less singular than $u'''$, the term $K$ is regular and cannot contribute, and so the dominant behaviour is governed by $\nu u''' + u^2/2$, which contributes the terms

$$\nu p(p-1)(p-2)u_0\chi^{p-3} + (1/2)(u_0\chi^p)^2, \tag{7}$$

and generates the two conditions

$$p - 3 = 2p, \ \nu p(p-1)(p-2)u_0 + (1/2)u_0^2 = 0, \ u_0 \neq 0. \tag{8}$$

Their unique solution is $p = -3, u_0 = 120\nu$, and the common value of the two powers is $q = p - 3 = 2p = -6$.

In a second step, one builds the linearized equation near the solution which behaves like $u_0\chi^p$

$$(\nu\partial_x^3 + u_0\chi^p)w = 0. \tag{9}$$

This linear ODE for $w$ has a Fuchsian singularity near $\chi = 0$ (see the textbook [11]), and one then computes its Fuchs indices, i. e. the roots $j$ of the algebraic equation

$$\lim_{\chi \to 0} \chi^{-j-q}(\nu\partial_x^3 + u_0\chi^p)\chi^{j+p} \tag{10}$$

$$= \nu(j-3)(j-4)(j-5) + 120\nu = \nu(j+1)(j^2 - 13j + 60) \tag{11}$$

$$= \nu(j+1)\left(j - \frac{13+i\sqrt{71}}{2}\right)\left(j - \frac{13-i\sqrt{71}}{2}\right) = 0. \tag{12}$$

A necessary condition for analyticity is that the two irrational Fuchs indices do not contribute, and this restricts to $3 - 2 = 1$ the number of arbitrary constants which the general analytic solution may depend on.

Let us now compute the similar quantities $u_0, p$ and the set of Fuchs indices for the (much more difficult) coupled CGL3 system.

## 4. Leading order of two coupled CGL3

One easily checks that $|A|$ and $|B|$ generically behave like simple poles. Let us then denote the dominant behaviour of the four fields $(A, \overline{A}, B, \overline{B})$ as [12]

$$A \sim a_0\chi^{-1+i\alpha}, \ \overline{A} \sim \overline{a}_0\chi^{-1-i\alpha}, \ B \sim b_0\chi^{-1+i\beta}, \ \overline{B} \sim \overline{b}_0\chi^{-1-i\beta}, \tag{13}$$

in which $(a_0, b_0)$ are complex constants, $(\alpha, \beta)$ real constants, all of them to be determined. The analogue of $p - 3 = 2p$ is now just the identity, and the analogue of the second equation in (8) is the set of four equations (the time-derivative term and the $\gamma$ term are not dominant)

$$p(-1+i\alpha)(-2+i\alpha) + q(a_2 + \delta b_2) = 0, \tag{14}$$

$$\overline{p}(-1-i\alpha)(-2-i\alpha) + \overline{q}(a_2 + \overline{\delta}b_2) = 0, \tag{15}$$

$$p(-1+i\beta)(-2+i\beta) + q(b_2 + \delta a_2) = 0, \tag{16}$$

$$\overline{p}(-1-i\beta)(-2-i\beta) + \overline{q}(b_2 + \overline{\delta}a_2) = 0. \tag{17}$$

with $a_2 = |a_0|^2, b_2 = |b_0|^2$. These are four nonlinear algebraic equations. The methodological error *not* to be done would be to solve them for the four real unknowns $(a_2, b_2, \alpha, \beta)$ as real expressions in the three complex parameters $(p, q, \delta)$, for this would generate algebraic (i. e. not rational) expressions and mess up all subsequent computations. On the contrary, one should make no distinction between the unknowns and the parameters, and consider this apparently nonlinear system as a *linear* system on the complex field $\mathcal{C}$ for a carefully chosen subset of four unknowns among the ten equivalent variables $(a_2, b_2, i\alpha, i\beta, p, q, \overline{p}, \overline{q}, \delta, \overline{\delta})$ so as to avoid the introduction of any algebraic quantity. Such a subset is $(a_2, b_2, \overline{p}, \overline{q})$. First, the subsystem (14), (16), linear inhomogeneous in $(a_2, b_2)$, is solved on $\mathcal{C}$ as

$$a_2 = \frac{p}{q(1-\delta^2)}\left(-(1-i\alpha)(2-i\alpha) + \delta(1-i\beta)(2-i\beta)\right), \tag{18}$$

$$b_2 = \frac{p}{q(1-\delta^2)}\left(-(1-i\beta)(2-i\beta) + \delta(1-i\alpha)(2-i\alpha)\right), \tag{19}$$

if one excludes the nongeneric case $\delta^2 = 1$, which is left to the reader. Second, the subsystem (15), (17) is linear homogeneous in $(\overline{p}, \overline{q})$, therefore its Jacobian must vanish (from now on, one restricts to $\overline{\delta} = \delta$)

$$\frac{q}{p}\frac{\mathrm{D}((15),(17))}{\mathrm{D}(\overline{p},\overline{q})} = (\alpha - \beta)(\alpha\beta + 2) = 0, \tag{20}$$

and the subsystem is solved on $\mathcal{C}$ as

$$\overline{p} = Kp(1-i\alpha)(2-i\alpha), \ \overline{q} = Kq(1+i\alpha)(2+i\alpha), \tag{21}$$

in which $K$ is an irrelevant arbitrary nonzero complex constant.

The constraint (20) between the real variables $(\alpha, \beta)$, which describe the argument of $(A, B)$, defines two mutually exclusive cases

1. $\alpha = \beta$, *a priori* not so interesting since it implies $a_2 = b_2$ and contains the reduction $A = B$ to one CGL3 equation,
2. $\alpha\beta = -2$, which forbids $\alpha = \beta$ and describes a truly coupled situation.

For reference, the resolution on $\mathcal{R}$ provides the algebraic solution

$$a_2 = \frac{9|p|^2}{2|q|^2(1-\delta^2)d_i^2}[(1-\delta)d_r + (1-\varepsilon\delta)\Delta], \tag{22}$$

$$b_2 = \frac{9|p|^2}{2|q|^2(1-\delta^2)d_i^2}[(1-\delta)d_r + (\varepsilon - \delta)\Delta], \tag{23}$$

$$\alpha = \frac{3}{2d_i}(d_r + \Delta), \ \beta = \frac{3}{2d_i}(d_r + \varepsilon\Delta), \tag{24}$$

$$\frac{p}{q} = d_r - id_i, \ \Delta^2 = d_r^2 + (8/9)d_i^2, \ \varepsilon^2 = 1, \tag{25}$$

which defines four *families of movable singularities* [10] for the dominant behaviour.

## 5. Fuchs indices

The indicial equation is now the determinant [10] of the fourth order matrix

$$\mathbf{P}(j) = \begin{pmatrix} (2a_0\overline{a_0} + \delta b_0\overline{b_0})q & qa_0^2 & \delta qa_0\overline{b_0} & \delta qa_0 b_0 \\ \overline{q}\overline{a_0}^2 & (2a_0\overline{a_0} + \delta b_0\overline{b_0})\overline{q} & \delta\overline{q}\overline{a_0}\overline{b_0} & \delta\overline{q}\overline{a_0}b_0 \\ \delta q\overline{a_0}b_0 & \delta qa_0 b_0 & (2b_0\overline{b_0} + \delta a_0\overline{a_0})q & qb_0^2 \\ \delta\overline{q}\overline{a_0}\overline{b_0} & \delta\overline{q}a_0\overline{b_0} & \overline{q}\overline{b_0}^2 & (2b_0\overline{b_0} + \delta a_0\overline{a_0})\overline{q} \end{pmatrix}$$
$$+ \mathrm{diag}(p(j-1+i\alpha)(j-2+i\alpha), \overline{p}(j-1-i\alpha)(j-2-i\alpha),$$
$$p(j-1+i\beta)(j-2+i\beta), \overline{p}(j-1-i\beta)(j-2-i\beta)). \tag{26}$$

In the limit $\alpha = \beta = 0$, it admits the symmetry $\forall j \ \mathbf{P}(j) = \mathbf{P}(3-j)$. After substitution of the values of $(a_2, b_2, \overline{p}, \overline{q})$ as given by (18), (19), (21), the variables $(p, q, K)$ are factored out and the determinant is a polynomial of degree eight in $j$ with coefficients polynomial in the real variables $(\alpha, \beta, \delta)$. Three of its zeros $j$ are already known : $-1$ as always in this kind of indicial equation, 0 twice, which correspond to the phase invariance of $A$ and $B$. Our interest here is to count the number of irrational indices, as done in Section 3. Let us split the discussion according to the two subcases (20).

When $\alpha = \beta$, the indicial equation factorizes as

$$\det \mathbf{P}(j) = (j+1)(j^2 - 7j + 6\alpha^2 + 12)j^2 \times \tag{27}$$
$$[(j+1)(j^2 - 7j + 6\alpha^2 + 12)(1+\delta) + 4\delta((2-\alpha^2)j - 3(2+\alpha^2))]. \tag{28}$$

If the coupling $\delta$ vanishes, one recovers as expected the square of the indicial equation of one CGL3 [13]. For generic values of the fixed parameters $(p, q, \delta)$, all indices, excepted $(-1, 0, 0)$, are irrational.

When $\alpha\beta = -2$, the remaining fifth degree polynomial does not factorize, and the indicial equation is

$$\det \mathbf{P}(j) = (j+1)j^2 \left[(1-\delta^2)(j^5 - 13j^4) + 40 \text{ terms}\right] = 0. \tag{29}$$

Again, there generically exist five irrational indices. Some particular values are

$$\alpha = -\beta, \ \det = (j+1)j^2(j^2 - 7j + 24)((1-\delta)(j^3 - 6j^2 + 17j) + 24(1+\delta)), \tag{30}$$
$$\delta = 1, \ \det = (j+1)j^2(\alpha^2 j^2 - (4 + 3\alpha^2 + \alpha^4)j + 3(2+\alpha^2)^2), \tag{31}$$
$$\delta = -1, \ \det = (j+1)j^2(\alpha^2(j^3 - 7j^2) + 3(2+\alpha^2)^2(j-3)). \tag{32}$$

## 6. Counting of the free constants

Let us count the precise number of arbitrary constants the general analytic solution for the solitary wave reduction (4) may depend on. The reduction introduces the three arbitrary constants $(c, \omega_A, \omega_B)$ and does not change the differential order eight of the PDE system. From the eleven possible arbitrary constants, one must subtract

1. the irrelevant origin $\xi_0$ of $\xi$ (Fuchs index $-1$), which represents the invariance under a space translation,
2. the irrelevant origin $\varphi_0$ of each phase (Fuchs indices $(0,0)$), which represents the invariance under a phase shift,
3. and the number of irrational Fuchs indices, generically five.

The result of this counting is summarized in Table 1.

Table 1: Maximal number of arbitrary constants in the solitary wave of one and two CGL3 equations.

| CGL3 | Reduction var | Diff. order | Irrelevant | Irrational Fuchs indices | Max arbitrary |
|---|---|---|---|---|---|
| 1 | $c, \omega$ | 4 | $\xi_0, \varphi_0$ | 2 | 2 |
| 2 | $c, \omega_A, \omega_B$ | 8 | $\xi_0, \varphi_{A,0}, \varphi_{B,0}$ | 5 | 3 |

Another counting, for the various possible topological structures, is made in Ref. [14].

## 7. Hints from the integrable limit

In order to have an idea of the class of expressions among which to search for exact solutions, let us consider the integrable limit of our system, i. e. the unique case $p$ real, $q$ real, $\gamma = 0$ and, for two coupled equations, $v = 0, \delta = \pm 1$ [15, 16].

For one equation, this is the famous nonlinear Schrödinger equation. The indices $(-1, 0, 3, 4)$ are all integral, and the general solution for the solitary wave

$$|A|^2 = a_2(\wp(\xi) - e_0), \tag{33}$$

$$\arg A = \omega t + \frac{c}{2p_r}\xi + (1/2)\sqrt{-(4e_0^3 - g_2 e_0 - g_3)} \int \frac{\mathrm{d}\xi}{\wp(\xi) - e_0}, \tag{34}$$

$$\omega = p_r((c/(2p_r))^2 + 3e_0), \tag{35}$$

in which $\wp$ is the Weierstrass elliptic function, depends on the four arbitrary constants $(c, \omega, g_2, g_3)$.

For two equations, the eight Fuchs indices $(-1, 0, 0, 0, 3, 3, 3, 4)$ are integral and the general solution for the solitary wave, which depends on eight arbitrary constants, is not yet known. Recently Porubov and Parker [17] found a six-parameter elliptic solution

$$\begin{cases} |A|^2 = a_2(\wp(\xi) - e_1), \\ |B|^2 = b_2(\wp(\xi) - e_2), \end{cases} \tag{36}$$

in which $(a_2, b_2)$ are those of Section 4., $e_1, e_2$ are constant, and also, because $\delta^2 = 1$, a five-parameter elliptic solution

$$\begin{cases} |A|^2 = A_1(\wp^2(\xi) + B_1\wp(\xi) + C_1), \\ |B|^2 = A_2(\wp^2(\xi) + B_2\wp(\xi) + C_2), \end{cases} \tag{37}$$

in which $A_k, B_k, C_k$ are constant.

When the elliptic function $\wp$ degenerates to a trigonometric function, the six-parameter solution (36) includes all the various front-type or pulse-type solitary waves [18, 16].

## 8. What was known for one and two CGL3

For one CGL3, one knows only four particular solutions of the reduction $\xi = x - ct$ with a zero codimension, i. e. without constraint on the fixed parameters $(p, q, \gamma)$. In all of them $|A|^2$ is a degree-two polynomial in $\tanh k\xi$, i. e. a trigonometric degeneracy of the integrable case (33). These are qualitatively ($K$ or $K_j$ denotes a real constant, and for brevity we omit the argument $k\xi$ of the functions)

1. a pulse or solitary wave [19]
$$|A|^2 = a_2 \operatorname{sech}^2, \ c = 0, \tag{38}$$

2. a front or shock [20]
$$|A|^2 = a_2(\tanh \pm 1)^2, \tag{39}$$

3. a source or propagating hole [21]
$$|A|^2 = a_2\left((\tanh + K_1)^2 + K_2^2\right), \ c \text{ arbitrary}, \tag{40}$$

4. an unbounded solution [22]
$$|A|^2 = a_2(\tan^2 + K^2), \ c = 0. \tag{41}$$

Only the propagating hole depends on one arbitrary constant (the velocity $c$). The missing solution should be, as indicated in Table 1 and explained in detail in Ref. [22], an extrapolation of this propagating hole to one more arbitrary constant.

Several other solutions have been "observed" in numerical simulations or in analytic perturbations and quite certainly correspond to unknown analytic solutions, these are

1. a propagating pulse [23], i. e. some extrapolation of (38) to an arbitrary velocity,
2. a homoclinic propagating hole [24], i. e. a solution with the shape of (40) but with the same limit of $|A|$ at both infinities $\xi \to \pm\infty$,
3. various other topological structures [25].

For two coupled CGL3 equations, the only analytic attempt we are aware of [26] results into equal phases for $A$ and $B$, i. e. essentially into the uncoupled situation. The result to be found should, according to Table 1, depend on three arbitrary constants.

## 9. Search for singlevalued global solutions

The difficulty comes from the fact that the natural physical variables $(A, \overline{A})$ or $(\operatorname{Re} A, \operatorname{Im} A)$ (it is sufficient for the moment to consider one CGL3) are *always* locally multivalued [13], hence not adapted to the search for singlevalued exact solutions. A more detailed study [22] uncovers the best representation for this purpose, namely a *complex modulus* $Z$ and a real argument $\Theta$ uniquely defined by

$$A = Z e^{i\Theta},\ \overline{A} = \overline{Z} e^{-i\Theta}. \tag{42}$$

For each of the families enumerated in Section 4., if one excludes the contribution of the irrational Fuchs indices, the three fields $(Z, \overline{Z}, \operatorname{grad} \Theta)$ are locally singlevalued and they behave like simple poles.

The physical variables $(|A|^2, \operatorname{grad} \arg A)$ also have this nice property of being locally singlevalued (they respectively behave like a double pole and a simple pole), but they are not as elementary as $(Z, \overline{Z}, \operatorname{grad} \Theta)$.

In order to obtain singlevalued exact solutions, one looks for representations, if they exist, of these locally singlevalued fields by Laurent series which terminate, thus ensuring *ipso facto* their closed form. This is the famous "truncation method" initiated by Weiss, Tabor, and Carnevale [27], the latest version of which is detailed in Ref. [28].

For two coupled CGL3 equations, this method can be applied either to the two couples $(|A|^2, \operatorname{grad} \arg A)$ or to the two couples $(Z, \overline{Z}, \operatorname{grad} \Theta)$. The first set can in principle, despite the much more involved computations, yield more solutions than the second set, but it should be noted that, for one CGL3, *all* known solutions are found with the second set. In order to shorten computations, let us handle the two couples $(Z, \overline{Z}, \operatorname{grad} \Theta)$.

## 10. How to minimize the computations

Quite similarly to Section 4., the natural methodology (to solve for real variables as real expressions) is *not* the one to follow, let us explain it on the truncation which provides the propagating hole solution (40) of one CGL3, since the two–CGL3 case will follow the same method.

The assumption is [22]

$$\begin{cases} Z = a_0(\chi^{-1} + X + iY), \\ \overline{Z} = \overline{a}_0(\chi^{-1} + X - iY), \\ \Theta = \omega t + \alpha \operatorname{Log} \psi + K\xi, \\ (\operatorname{Log} \psi)' = \chi^{-1},\ \chi' = 1 - (k^2/4)\chi^2, \end{cases} \tag{43}$$

in which $\chi$ and $\psi$ are functions of $\xi = x - ct$, $(a_0, \alpha)$ are the constants of Section 4., $(\omega, X, Y, K, k^2)$ are real constants. After elimination of any derivative of $\operatorname{Log} \psi$ and $\chi$, the lhs $E$ of the CGL3 becomes a Laurent series which terminates

$$E e^{-i\Theta} = \sum_{j=0}^{3} E_j \chi^{j-3}, \tag{44}$$

and one has to solve the four complex (eight real) equations $E_j = 0$ in the eight real unknowns $(a_2, \alpha, \omega, X, Y, K, c, k^2)$, the two complex parameters $(p, q)$, and the real parameter $\gamma$. If there exists a solution, the elementary building block functions evaluate to

$$\chi = \frac{k}{2} \tanh \frac{k\xi}{2}, \ \psi = \cosh \frac{k\xi}{2}. \tag{45}$$

The good methodology is again to select, among the eleven complex variables considered as equivalent, four variables which make the system a *linear* one of Cramer type. In this example, one proceeds as follows [22]. The system $(E_0, E_1, E_2)$ is of Cramer type in $(a_2, K, \omega)$, and after its resolution the last equation $E_3$ is independent of $(p, q, \gamma, c)$ and factorizes into a product of linear factors

$$E_3 \equiv [k^2 - 4(X + iY)^2](\alpha Y - 2X) = 0. \tag{46}$$

Let us apply this to the coupled CGL3 system.

## 11. Double holes

The analogue of the hole solution of one CGL3 [21] is searched with the assumption

$$\begin{cases} A = a_0(\chi^{-1} + X_a + iY_a)e^{+i(\omega_a t + \alpha \log \psi + K_a \xi)}, \\ \overline{A} = \overline{a}_0(\chi^{-1} + X_a - iY_a)e^{-i(\omega_a t + \alpha \log \psi + K_a \xi)}, \\ B = b_0(\chi^{-1} + X_b + iY_b)e^{+i(\omega_b t + \beta \log \psi + K_b \xi)}, \\ \overline{B} = \overline{b}_0(\chi^{-1} + X_b - iY_b)e^{-i(\omega_b t + \beta \log \psi + K_b \xi)}, \end{cases} \tag{47}$$

with the same definition for $(\log \psi)'$ and $\chi'$ as in previous Section. These are eight complex equations in fourteen real unknowns $(a_2, b_2, \alpha, \beta, \omega_a, \omega_b, X_a, Y_a, X_b, Y_b, K_a, K_b, c, k^2)$, two complex parameters $(p, q)$, and three real parameters $(v, \gamma, \delta)$. After solving the six equations $(E_0, E_1, E_2)$ as a Cramer system on $\mathcal{C}$ in $(a_2, b_2, K_a, K_b, \omega_a, \omega_b)$, the last two equations $E_3$ factorize into a product of linear equations:

$$\left[k^2 - 4(X_a + iY_a)^2\right] F(Y_a, X_a, X_b, \alpha, \beta) = 0, \tag{48}$$

$$\left[k^2 - 4(X_b + iY_b)^2\right] F(Y_b, X_b, X_a, \beta, \alpha) = 0, \tag{49}$$

$$F(Y, X_a, X_b, \alpha, \beta) \equiv \alpha Y - (X_a + X_b)$$

$$- (X_a - X_b) \frac{1 - i\alpha - \delta(1 - i\beta) + \delta^2(1 - i\alpha)^2}{(1 - \delta^2)(1 - i\alpha)}. \tag{50}$$

The bifurcations successively encountered during the resolution on $\mathcal{R}$ are

$$j = 0: \quad (\beta = \alpha) \text{ or } (\alpha\beta = -2), \tag{51}$$

$$j = 1: \quad p_i \text{ zero or nonzero}, \tag{52}$$

$$j = 2: \quad (\alpha - \beta)p_r \text{ zero or nonzero}, \tag{53}$$

$$j = 3: \quad \text{the three subcases of (48)–(49).} \tag{54}$$

The solutions $p_r p_i \neq 0$ which do not reduce to a mere product of two CGL are

$$X_b = X_a, \ \alpha Y_a = \beta Y_b = 2X_a, \ v = 0, \ \alpha\beta = -2, \ c = 6p_i^{-1}|p|^2 X_a,$$
$$k^2 = -4(3 + (\alpha+\beta)p_i/p_r)(3 + (\alpha+\beta)^2\delta/(1+\delta))X_a^2, \tag{55}$$
$$X_b = X_a, \ Y_a = Y_b = 0, \ v = 0, \ k^2 = 4X_a^2, \ \alpha\beta = -2, \ c = 6p_i^{-1}|p|^2 X_a, \tag{56}$$
$$X_b = -X_a, \ Y_a = Y_b = 0, \ c = 0, \ k^2 = 4X_a^2, \ \alpha = \beta \tag{57}$$
$$X_b = -X_a, \ Y_a = Y_b = 0, \ k^2 = 4X_a^2, \ \alpha\beta = -2,$$
$$c = -2(\alpha^2 - \beta^2)p_i^{-1}|p|^2 X_a \delta/(1-\delta^2),$$
$$v = -2((1+\delta)(3+\delta) + (\alpha+\beta)^2 \delta)p_i^{-1}|p|^2 X_a/(1-\delta^2), \tag{58}$$
$$X_b = X_a, \ Y_a = 0, \beta Y_b = 2X_a, \ v = 0, \ c = 6p_i^{-1}|p|^2 X_a, \ k^2 = 4X_a^2, \tag{59}$$
$$Y_a = 0, \beta Y_b = 2X_b, \ k^2 = 4X_a^2, \ \alpha\beta = -2,$$
$$c = (X_a + 5X_b)|p|^2/p_i, v = (X_b - X_a)|p|^2/p_i, \tag{60}$$

with the respective codimensions $1, 1, 1, 2, 2, 2$. The fixed constraints are respectively

$$v = 0, \tag{61}$$
$$v = 0, \tag{62}$$
$$|p|^2(3-\delta)^2\gamma - P_1(\delta, p_r, p_i, \alpha)p_i v^2 = 0, \tag{63}$$
$$|p|^2((1+\delta)(3+\delta) + (\alpha+\beta)^2\delta)^2\gamma - P_2(\delta, p_r, p_i, \alpha+\beta)p_i v^2 = 0,$$
$$P_3(\delta, p_r, p_i, \alpha+\beta) = 0, \tag{64}$$
$$v = 0, \ P_4(\delta, p_r, p_i, \alpha, \beta) = 0, \tag{65}$$
$$\alpha^2 + 2\delta = 0, \ P_5(p_r, p_i, \alpha)\gamma^2 + P_6(p_r, p_i, \alpha)\gamma v^2 + P_7(p_r, p_i, \alpha)v^4 = 0, \tag{66}$$

in which $P_n$ denotes a polynomial of its arguments.

These "double hole" solutions depend on no arbitrary at all (as compared to the maximal value of three, see Section 6.), they are all heteroclinic and can never be homoclinic.

## 12. Double pulse solution

A double pulse similar to (38) is searched with the assumption

$$\begin{cases} A = \{a_0(\chi^{-1} + X_a + iY_a) + (k^2/4)a_1\chi\} e^{+i(\omega_a t + \varphi_A + K_a \xi)}, \\ \overline{A} = \{\overline{a}_0(\chi^{-1} + X_a - iY_a) + (k^2/4)\overline{a}_1\chi\} e^{-i(\omega_a t + \varphi_A + K_a \xi)}, \\ B = \{b_0(\chi^{-1} + X_b + iY_b) + (k^2/4)b_1\chi\} e^{+i(\omega_b t + \varphi_B + K_b \xi)}, \\ \overline{B} = \{\overline{b}_0(\chi^{-1} + X_b - iY_b) + (k^2/4)\overline{b}_1\chi\} e^{-i(\omega_b t + \varphi_B + K_b \xi)}, \\ \varphi'_A = \alpha\chi^{-1} + (k^2/4)\alpha_1\chi, \ \varphi'_B = \beta\chi^{-1} + (k^2/4)\beta_1\chi, \end{cases} \tag{67}$$

in which $a_0$ and $a_1$ are two different roots of the leading order equation, so as to produce a sech according to the elementary identities [22]

$$\tanh z - \frac{1}{\tanh z} = -2i \operatorname{sech}\left[2z + i\frac{\pi}{2}\right], \ \tanh z + \frac{1}{\tanh z} = 2\tanh\left[2z + i\frac{\pi}{2}\right]. \tag{68}$$

This generates fourteen complex equations in the eighteen real unknowns
$(a_0, b_0, a_1, b_1, \alpha, \beta, \alpha_1, \beta_1, \omega_a, \omega_b, X_a, Y_a, X_b, Y_b, K_a, K_b, c, k^2)$, the two complex parameters $(p, q)$, and the three real parameters $(v, \gamma, \delta)$. Among the eight cases of signs in (24), let us restrict to the two which allow $a_0^2 = a_1^2$ and $b_0^2 = b_1^2$; their sech-type solution is

$$a_1 = -a_0, \ |a_0|^2 = a_2, \ \alpha_1 = \alpha, \ X_a = 0, \ Y_a = 0, \ K_a = (c-v)p_r/(2|p|^2), \quad (69)$$
$$b_1 = -b_0, \ |b_0|^2 = b_2, \ \beta_1 = \beta, \ X_b = 0, \ Y_b = 0, \ K_b = (c+v)p_r/(2|p|^2), \quad (70)$$
$$cp_i = 0, \ vp_i = 0, \ (\alpha - \beta)[(\alpha + \beta)p_i + 2p_r] = 0, \quad (71)$$
$$\gamma = ((1 - (\alpha^2 + \beta^2)/2)p_i - (\alpha + \beta)p_r)k^2. \quad (72)$$

The two moduli are proportional to sech $kx$ but the two phases are different if $\alpha \neq \beta$. The codimension is either one (case $\alpha = \beta$) or two (case $\alpha\beta = -2$), and, if $p_i = 0$, the two velocities are arbitrary.

### 13. Conclusion

From the singularity structure, we have found several exact solutions to the coupled CGL3 system. At the time being, these are the only known analytic solutions to this system. Their main features are

1. their physical relevance, since they represent structures observed in experiments [6, 7],
2. their hole or bell-shaped profile characterizing their nonlinearity (these are not plane waves),
3. their true coupling.

However, a drawback is their nonzero codimension.

In particular, Bekki–Nozaki holes have been experimentally observed in a rectangular geometry with a nonzero group velocity $v$ [2].

The gap between the found results and the expected results described in Section 6. is much more important for the two–CGL3 system (three missing arbitrary constants, nonzero codimension) than for one CGL3 (one missing arbitrary constant, zero codimension). Therefore the effort should be put on filling the one–CGL3 gap, before tackling again the two–CGL3 system.

### 14. Acknowledgements

RC thanks the University of Toruń for invitation and the Ministère des affaires étrangères for travel support. Both authors acknowledge the financial support of the Tournesol grant No. T99/040. MM acknowledges the financial support of the IUAP Contract No. P4/08 funded by the Belgian government and the support of CEA.